\documentclass[
 reprint,
 amsmath,amssymb,
 aps,
prb,
floatfix,
]{revtex4-2}

\usepackage{amsmath}
\usepackage{amssymb} 
\usepackage{amsthm}
\usepackage[english]{babel}
\usepackage{bm}
\usepackage{braket}
\usepackage{dcolumn}
\usepackage{float}
\usepackage[bottom]{footmisc} 
\usepackage{graphicx}
\usepackage{hyperref}
\usepackage{lipsum}
\usepackage{physics}
\usepackage{placeins}
\usepackage{textcomp, mathcomp}
\usepackage{xcolor}
\usepackage{times}

\newcommand{\supplementarysection}{%
  \setcounter{figure}{0}
  \let\oldthefigure\thefigure
  \renewcommand{\thefigure}{S\oldthefigure}
  \section*{Supplemental material}
  }
  
\newcommand{\bst}{(Bi$_{1-x}$Sb$_x$)$_2$Te$_3$ }
\newcommand{\bsttight}{(Bi$_{1-x}$Sb$_x$)$_2$Te$_3$}

\begin{document}

\preprint{APS/123-QED}

\title{\Large Signature of current-induced nuclear spin polarization in (Bi$_{1-x}$Sb$_{x}$)$_2$Te$_3$}

\author{Sofie Kölling\textsuperscript{1}}
\author{\.{I}nan\c{c} Adagideli\textsuperscript{1, 2}}
\author{Alexander Brinkman\textsuperscript{1}}

\affiliation{\textsuperscript{1}MESA+ Institute for Nanotechnology, University of Twente, The Netherlands}
\affiliation{\textsuperscript{2}Faculty of Engineering and Natural Sciences, Sabanci University, Turkey}

\date{\today}
\begin{abstract}
In systems with spin-momentum locking, such as the surface states of three-dimensional topological insulators, a charge current is spin-polarized and spin-flip interactions between electron and nuclear spins can transfer this polarization to the nuclear spin system. When a nonzero bias voltage is applied, the nuclear polarization reaches a steady-state value. This polarization emerges as an effective in-plane magnetic field acting on electrons, called the Overhauser field, which causes an offset in-plane magnetoresistance perpendicular to the current, visible in experiments. The in-plane offset is measured in the three-dimensional topological insulator \bsttight, and the magnitude of the magnetic field offset is compared to the Overhauser field. We attribute the observed magnetic field offset to current-induced nuclear polarization in \bsttight, which forms an important step towards experimentally realizing an entropic inductor.
\end{abstract}
\maketitle

Developments in electronic materials lead to novel methods of energy storage based on information. As first discussed in a thought experiment known as `Maxwell's Demon' \cite{knott1911life}, work can be extracted from a system by utilizing the information contained within. Such processes satisfy the second law of thermodynamics due to Landauer's principle \cite{landauer1961irreversibility}, which states that heat is dissipated upon erasing information, required to reset the engine cycle. Accordingly, the upper limit of extractable work equals $k_\mathrm{B}T\ln{2}$ per bit. While this scalable limit holds potential for extracting substantial amounts of work, most experimental realizations of Maxwell's demon rely on feedback-based systems \cite{koski2015chip, vidrighin2016photonic, toyabe2010experimental}, in which scalability remains a challenge.

The surface states of three-dimensional topological insulators are a suitable platform to resolve this challenge, by utilizing information encoded in nuclear spins \cite{bozkurt2018Work, bozkurt2024entropy}.
Through hyperfine interaction, a spin-momentum locked current can transfer electron spin polarization to a nuclear spin polarization via spin-flip interactions. Reversedly, thermal relaxation of the polarized nuclear spins would induce a charge current, driven by the information entropy change of the nuclear spin system and thereby converting heat into electric work. The entropy-driven current is inductive in nature, and proportional to the number of nuclear spins, emphasizing the scalability of the system \cite{bozkurt2024entropy}. 
Beyond the theoretical description, such an entropic inductor has not yet been experimentally realized, and confirming whether the nuclear spins are polarized by a spin-momentum locked current is a crucial first step.

Achieving this effect requires a nuclear spin-abundant material with spin-momentum locking. This requirement could be fulfilled by the three-dimensional topological insulator \bst (BST) \cite{zhang2009topological}, as the constituent isotopes have finite nuclear spin ($I_\mathrm{Bi} = 9/2$, 100\% abundant, $I_\mathrm{Sb} = 5/2$ or $7/2$, 57\% and 43\% abundant respectively, and $I_\mathrm{Te} = 1/2$, 8\% abundant \cite{schliemann2003electron}). Furthermore, the position of the Fermi energy and Dirac cone with respect to conduction and valence bands can be engineered \cite{zhang2011band}, reducing bulk conductivity.
Nuclear spin polarization has been studied in transport experiments on other bismuth-based materials. In the Rashba system Bi(111), a finite nuclear spin polarization was measured through a suppression of weak antilocalization (WAL) \cite{jiang2020dynamic}, and spin potentiometric measurements on Bi$_2$Te$_2$Se flakes show effects of a persistent nuclear spin polarization, after applying a large `writing current' \cite{tian2014quantum}. 

In this work, we investigate signatures of current-induced nuclear spin polarization in BST on timescales far exceeding the average nuclear polarization timescale $\tau_m$ \footnote{For BST, $\tau_m \sim 10^{-1} - 10^{1}$ seconds at $T = 4.5$ K (for an estimate, see \cite{bozkurt2024entropy}). Unfortunately, this timescale is relatively long, which limits the magnitude of the induced current, requiring highly detailed measurements to resolve the inductive effect.}. On these extended timescales, the nuclear spin-flip rate approaches zero, resulting in Ohmic (non-inductive) conduction according to theory \cite{bozkurt2024entropy}. However, this steady-state differs from a system without nuclear spin; at finite bias voltages, the mean nuclear polarization $m$ will attain a nonzero steady-state value, balancing polarization due to spin-momentum locked electrons and thermal relaxation.
This nonzero steady-state nuclear/magnetic polarization can affect transport through other mechanisms, not considered in previous models of an entropic inductor \cite{bozkurt2024entropy}, which we will study here.

\begin{figure*}
    \centering
    \includegraphics[width=\textwidth]{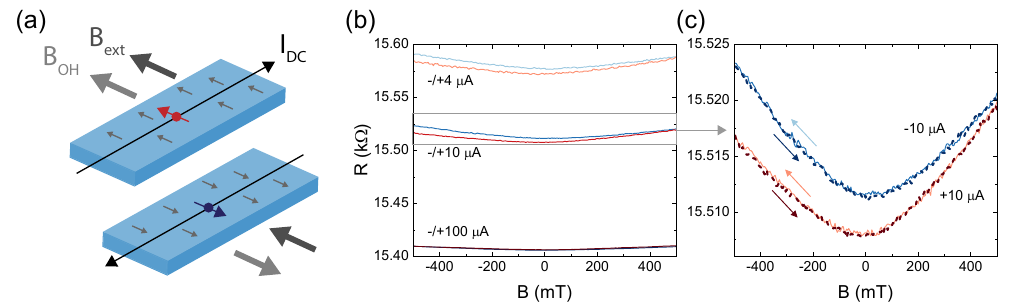}
    \makeatletter\long\def\@ifdim#1#2#3{#2}\makeatother
    \caption{(a) Measurement geometry where $I_\mathrm{DC}$ causes a nonequilibrium nuclear polarization. The resultant Overhauser field adds to the external magnetic field, causing an offset. (b) Measured $R_{xx}$ in a \bst Hall bar at $T = 4.5$ K with $W\times L = 6\ \mu$m$\times 60\ \mu$m as a function of in-plane magnetic field, for a range of $I_\mathrm{DC}$.  (c) Zoomed-in data showing the traces measured at $I_\mathrm{DC} =  \pm 10\ \mu$A. Both magnetic field sweep directions (up/down) are plotted, which have been measured back-to-back. Any shift in the magnetoresistance is independent of magnetic field sweep direction, and occurs on timescales much shorter than the measurement time ($\sim$ minutes).}
    \label{fig:inplane_setup}
\end{figure*}

We present a direct method of probing the nuclear spin polarization by focusing on the Overhauser field, which is an effective magnetic field acting on electrons generated by the nuclear spins \cite{fleisher1984optical}. Probing nuclear polarization via the Overhauser field has proved successful in previous research by Jiang \textit{et al.} \cite{jiang2020dynamic}. However, whereas Jiang \textit{et al.} probe the Overhauser field via its effect on the dephasing timescale using an out-of-plane magnetic field, we will focus on an in-plane magnetic field, perpendicular to the current as depicted in Fig.~\ref{fig:inplane_setup}(a). This direction is collinear with the expected nuclear polarization. 

The Overhauser field adds to the Zeeman interaction introduced by this external magnetic field and thereby shifts the magnetoresistance along the magnetic field axis. 
Previous studies indicate that such an in-plane external magnetic field does not influence dynamic nuclear polarization \cite{jiang2020dynamic}. 
Consequently, in this configuration, the offset in magnetoresistance due to the Overhauser field is set by the spin-momentum locked current through the topological surface state. 

We measure the magnetoresistance in BST Hall bars in the in-plane configuration and compare the results to predictions based on literature. The magnitude of the offset matches the expected Overhauser field in BST, and is suppressed at high currents, which could be an effect of Joule heating \cite{nandi2018logarithmic, anderson1979possible, abrahams1980non}. We present a proof-of-principle for current-induced nuclear polarization in BST in an otherwise non-magnetic device, which takes us one step closer to experimentally realizing an entropic inductor.

\textit{Estimate of the Overhauser field --} Prior to discussing experiments, we estimate the Overhauser field in BST. Transport in topological surface states is modeled by the Hamiltonian $H_0 = \hbar v_F \left(\bm{k} \times \bm{\sigma}\right)\cdot \hat{z}$ \cite{zhang2009topological, fu2009hexagonal}, 
where $v_F$ is the Fermi velocity, $\bm{k} = \left(k_x, k_y\right)$ is the momentum operator, and $\bm{\sigma}$ is the vector of Pauli matrices describing the spin degree of freedom of the charge carriers.

For materials in the Bi$_2$Te$_3$ family, this Hamiltonian is incomplete, because the Fermi surface is hexagonally warped away from the Dirac point \cite{fu2009hexagonal}. This warping introduces an out-of-plane component to the spin polarization, and is described by $H_\mathrm{HW} = \lambda k^3 \cos{3\phi_{k}}\sigma_z$,
where $\lambda$ is the hexagonal warping strength and $\phi_k$ is the angle of $\bm{k}$ relative to the x-axis. 

An external magnetic field affects both the orbital and the spin terms in the Hamiltonian. The effect on the orbital terms is captured by writing the (static) external magnetic field as vector potential ($\bm{B} = \nabla \times \bm{A}$), and substituting $\bm{k} \rightarrow \bm{k} - \frac{q}{\hbar}\bm{A}$. The effect on the spin terms is described by the Zeeman interaction $H_\mathrm{Z} = g \mu_B \bm{B}\cdot \bm{\sigma}$.

Our goal is to measure a response from an Overhauser field, which can be generated via dynamic nuclear polarization in a topological surface state: a current is spin-polarized due to spin-momentum locking, and this spin polarization can be transferred to nuclear spins via spin-flip interactions. 
We estimate the Overhauser field in BST by following an analysis similar to Jiang \textit{et al.}, who considered nuclear spin polarization generated by the Rashba states on Bi(111) surfaces \cite{jiang2020dynamic}. We verify that the electron spin polarization is sufficient to dynamically polarize nuclear spins without external magnetic field \footnote{See Supplemental Material A at [URL will be inserted by publisher] for an analysis on achieving dynamic nuclear polarization. See also references \cite{salis2009signatures, mulder2022revisiting, kondou2016fermi, mulder2022spectroscopic} therein.}.\nocite{salis2009signatures, mulder2022revisiting, kondou2016fermi, mulder2022spectroscopic}

The Overhauser field can be modeled by considering the hyperfine interaction ($H_\mathrm{HF}$) as an effective Zeeman contribution \cite{dyakonov1984theory, fleisher1984optical}, using
\begin{equation}
    H_\mathrm{HF} = g \mu_B \bm{B}_\mathrm{OH}\cdot\bm{\sigma} = A_0 v_0 \sum_n \bm{I}^n \cdot \bm{\sigma} \delta(\bm{r} - \bm{r_n}),
\end{equation}
where $A_0$ is the hyperfine interaction energy, $v_0$ is the unit cell volume and $\bm{I}^n$ is the nuclear spin at position $\bm{r}_n$.
We replace the sum by an integral, and consider that the average volume per nuclear spin equals $v_0/[n]$, where $v_0$ is the unit cell volume and $[n]$ is the number of nuclei per unit cell. Here, we assume that the nuclear spin density is not yet incorporated in $A_0$, similar to the model described by D'yakonov \cite{dyakonov1984theory}.
Besides, we replace the nuclear spin polarization by the average value $\bm{I}_\mathrm{av}$, and find
\begin{equation}
    H_\mathrm{HF} = A_0 v_0 \int \bm{I}_\mathrm{av} \cdot \bm{\sigma} \delta(\bm{r})\frac{[n]}{v_0}\dd^3 \bm{r} = A_0[n]\bm{I}_\mathrm{av} \cdot \bm{\sigma}.
\end{equation}
For the surface state of a three-dimensional topological insulator, the electron spin polarization is perpendicular to the momentum, and the average nuclear polarization generated by spin-flip interactions will be collinear with the electron spin polarization. Therefore, we find that $\bm{B}_\mathrm{OH}$ is perpendicular to the current, and its magnitude equals
\begin{equation}\label{eq:Boh}
    B_\mathrm{OH} =[n] \frac{A_0 I_\mathrm{av}}{g \mu_B}. 
\end{equation}

We find $I_\mathrm{av}$ by calculating the current-induced nuclear polarization on the surface of a three-dimensional topological insulator, taking into account both nonmagnetic and magnetic (nuclear spin) scattering. For the full derivation, see \cite{bozkurt2024entropy}. Assuming that $I_\mathrm{av}$ equals its steady-state value,
\begin{equation}\label{eq:mbar}
    I_\mathrm{av} = I \tanh{\left( \frac{\ell_\mathrm{el}}{L}\frac{eV}{2k_\mathrm{B}T} \right)}.
\end{equation}

Using a sheet current density of $j = 1$ A/m in a typical device with $L\times W = 60 \times 6$ $\mu$m, $\ell_\mathrm{el} = 10$ nm, $R = 15$ k$\Omega$, a range of $A_0 = 5-50$ $\mu$eV \cite{bozkurt2024entropy} and $g = 30$ \cite{drath1967band} for BST subsequently results in $B_\mathrm{OH} = 2.2 - 22$ mT. In the high bias limit, the steady-state  $I_\mathrm{av} = I$. Then, we find $B_\mathrm{OH} = 0.058-0.58$ T.

The Overhauser field has been estimated for a single TI surface. In reality, both the top and bottom surfaces contribute. Spin-momentum locking is opposite on both surfaces, so the total effect will be reduced compared to a single surface, and the calculated $B_\mathrm{OH}$ is an upper limit to what the experiments will yield.

\textit{Experimental methods --} We fabricate a BST Hall bar to measure the effects of an Overhauser field. To this end, BST thin films with $x = 0.6$ and a thickness of approximately 10 nm are deposited on Al$_2$O$_3$ substrates by molecular beam epitaxy. The films are structured into Hall bars using argon milling, and Ohmic contacts are deposited by sputter deposition of tungsten. The devices are capped with AlO$_x$ via atomic layer deposition. 

The Hall bar measured in this work has $W = 6\ \mu$m and $L =60\ \mu$m, and is also studied in \cite{MR_app},\nocite{hikami1980spin, garate2012weak} where the out-of-plane magnetoresistance is discussed in detail. 
The measurements are performed in a physical properties measurement system (PPMS) at $T = 4.5$ K, with the external magnetic field perpendicular to the current and oriented in-plane as shown in Fig.~\ref{fig:inplane_setup}(a), and we measure the magnetoresistance for a range of $I_\mathrm{DC}$ up to 100 $\mu$A (or $j$ up to 17 A/m), thereby varying any current-induced nuclear polarization \footnote{The magnetoresistance has also been measured at low $I_\mathrm{DC}$ after applying $I_\mathrm{DC} = 100$ $\mu$A for elongated time periods up to 2 hours. However, this yielded no measurable change when comparing the magnetoresistance before and after charging. The timescale at which the results discussed in this chapter set in must therefore be smaller than the time it takes to obtain these measurements (5 minutes).}. 

\textit{In-plane magnetoresistance --} Figure~\ref{fig:inplane_setup}(b) shows the measured longitudinal resistance ($R_{xx}$) as a function of the in-plane magnetic field, for a range of $I_\mathrm{DC}$. We observe two effects: overall, $R_{xx}$ is suppressed when increasing $I_\mathrm{DC}$. We attribute this to Joule heating \cite{MR_app}. On top of the suppression, a clear asymmetry as function of $I_\mathrm{DC}$ is observed. Figure~\ref{fig:inplane_setup}(c) highlights the up- as well as down-sweep direction for $I_\mathrm{DC} = \pm10\ \mu$A, which emphasizes that the asymmetry is reproducible and not a time-dependent (hysteretic) measurement artifact.
Finding whether the origin of this asymmetry corresponds to the Overhauser field requires a model describing the in-plane magnetoresistance of a topological surface state, which we discuss next.

Without hexagonal warping, adding the Zeeman contribution to the system is equivalent to a uniform shift in the momenta and can therefore be removed by a gauge transformation. This transformation suggests that the scattering probabilities are unaffected by an in-plane magnetic field. 
Upon including hexagonal warping, the influence of an in-plane magnetic field on quantum corrections becomes apparent when we consider that the contribution from the in-plane field cannot be gauged away. 
The resulting Zeeman field breaks time-reversal symmetry, and introduces a mass term in the Hamiltonian, thereby suppressing weak antilocalization and causing a crossover to weak localization as modeled by Adroguer \textit{et al}. \cite{adroguer2012diffusion}. Because this effect concerns a Zeeman interaction, appearing in the Hamiltonian in the same way as the Overhauser field, the Overhauser field adds to this component of the magnetoresistance. The magnetic field is replaced by $B_\mathrm{eff} = B_\mathrm{ext} + B_\mathrm{offs}$, resulting in an offset
magnetoresistance as function of external parallel field. Here $B_\mathrm{offs}$ includes $B_\mathrm{OH}$ and other possible current-induced effective fields 
\cite{Bfields_app}.\nocite{balram2019current, lundgren2015landau, baum2012magnetic}

In the warping model considered by Adroguer \textit{et al.} \cite{adroguer2012diffusion}, an in-plane magnetic field acts as an additional magnetic dephasing with length scale $L_B = \sqrt{D\tau_e/m_\mathrm{HW}}$, where $D$ is the diffusion constant, $\tau_e$ is the momentum relaxation time and mass term $m_\mathrm{HW} \approx m_\mathrm{HW}^0\cdot(g\mu_BB_\mathrm{eff}/E_F)^2$ encoding the effects of the Zeeman field \cite{adroguer2012diffusion}. 
The total dephasing length ($\Tilde{L}$) consists of the magnetic dephasing length and the phase coherence length ($L_\phi$) as $\Tilde{L}^{-2} = L_B^{-2} + L_\phi^{-2}$, and the resulting quantum correction $\delta \sigma_{xx}(B)\propto \ln{(\Tilde{L}/\ell_\mathrm{el})}$. As a result, the relative magnetic field dependence of the conductivity from the quantum corrections becomes $\Delta \sigma_{xx} = \delta \sigma_{xx}(B_\mathrm{eff}) - \delta \sigma_{xx}(0)$ and equals \footnote{Conventionally, the quantum corrections are calculated with respect to $B_\mathrm{ext} = 0$. However, at $B_\mathrm{ext} = 0$ we have $B_\mathrm{eff} = B_\mathrm{offs}$, so we calculate the corrections relative to $B_\mathrm{eff} = 0$ instead. In experiments, we assume that $B_\mathrm{eff} = 0$ where $\sigma_{xx}$ reaches its maximum value, and use $\Delta \sigma_{xx} = \sigma_{xx} - \mathrm{max}(\sigma_{xx})$.}

\begin{equation}\label{eq:Bpar}
    \Delta \sigma_{xx}(B) = -\alpha\frac{e^2}{\pi h}\ln\left( 1 + \frac{L_\phi^2}{L_B^2}\right) = -\alpha\frac{e^2}{\pi h}\ln\left( 1 + \frac{B_\mathrm{eff}^2}{B_{c}^2}\right).
\end{equation}
Here, $B_{c} = (L_B/L_\phi)\cdot B_\mathrm{eff}$, and $\alpha$ is a proportionality factor. 

Apart from this contribution due to hexagonal warping, competing effects occur. 
Firstly, a residual out-of-plane component in $B_\mathrm{ext}$ (from experimental error) would add a symmetric contribution to the magnetoresistance.
Secondly, the surface state is not purely 2D, and the parallel magnetic field threads the finite penetration depth of the surface state as described by Tkachov and Hankiewicz \cite{tkachov2011weak}. This effect breaks down quantum corrections similar to a perpendicular magnetic field, although the characteristic lengths are rescaled due to the confinement of the surface state. The result is a suppression of WAL as well, with a similar functional form to Eq.~(\ref{eq:Bpar}) where $\alpha = 1/2$ is the expected prefactor for three-dimensional topological insulators \cite{tkachov2011weak}.  However, this model only concerns the \textit{orbital} terms in the Hamiltonian, to which the Overhauser field (a Zeeman term, acting on the spin) does not add. Therefore, the surface state threading described by the Tkachov-Hankiewicz model causes an additional contribution to the magnetoresistance, symmetric in $B_\mathrm{ext}$. 

Combining these considerations, although Eq. (\ref{eq:Bpar}) does not capture the complete physics of the quantum corrections in an in-plane magnetic field, it is the only term affected by the Overhauser field. Therefore, in the following section we will fit the data to Eq. (\ref{eq:Bpar}) using $B_\mathrm{eff}$ and relate the offset to the Overhauser field. The competing dephasing effects will likely cause a deviation from the fit.

\begin{figure}
    \centering
    \includegraphics{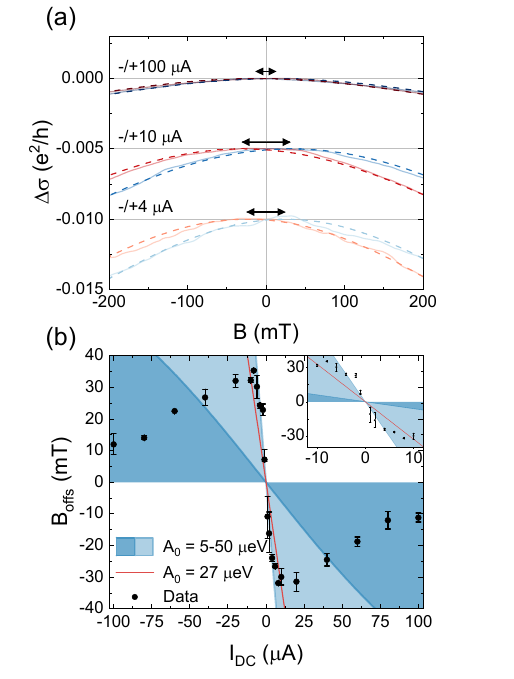}
    \caption{(a) $\Delta \sigma_{xx}(B) =  \sigma_{xx}(B) - \sigma_{xx}(0)$ calculated from the data in Fig.~\ref{fig:inplane_setup}(b) (solid lines) and the fitted Eq.~(\ref{eq:Bpar}) (dashed), at $T = 4.5 K$. Left/right arrows denote the negative/positive shift extracted from the fit. The traces are offset in $\Delta\sigma$ for visibility, grey horizontal lines denote the zero values. (b) Fit parameter $B_\mathrm{offset}$ as function of $I_\mathrm{DC}$, with an enlarged graph for small $I_\mathrm{DC}$ in the inset. The solid dots are the fit values obtained on an averaged dataset, and the error bars denote the minimum and maximum shift obtained when considering both sweep directions of $B_\mathrm{ext}$ (negative $\rightarrow$ positive and positive $\rightarrow$ negative). The light blue area is the calculated range of $B_\mathrm{OH}$ using a single BST topological surface, and the dark blue area denotes the range below the minimum calculated $B_\mathrm{OH}$, in which the measured $B_\mathrm{OH}$ could fall due to partial cancellation from the top and bottom surface nuclear polarization. The red line corresponds to the value of $A_0 = 27$ $\mu$eV found by Nisson \textit{et al.} \cite{nisson2013nuclear} using NMR at $T = 10$ K}.
    \label{fig:inplane_results}
   
\end{figure}

\textit{Extracting the offset --} We calculate $\sigma_{xx} = (L/W)/R_{xx}$ and subsequently $\Delta \sigma_{xx}(B) = \sigma_{xx}(B) - \sigma_{xx}(0)$ which is shown in Fig.~\ref{fig:inplane_results}(a). $\sigma_{xx}$ is suppressed with increasing magnetic field, implying a WAL suppression.
Equation~(\ref{eq:Bpar}) is fitted to $\Delta \sigma_{xx}$, with fit parameters $\alpha$, $B_c$, and an offset in $B$. This equation captures the suppression of quantum corrections due to Zeeman-like effects, as function of the effective in-plane magnetic field ($B_\mathrm{eff} = B_\mathrm{ext} + B_\mathrm{offs}$). 

Fitting the data requires $\alpha = 0.01$ at high current and $\alpha = 0.05$ at low current. This low value of $\alpha$ could indicate that a transport channel with weak localization is present as well, or other effects not captured by Eq.~(\ref{eq:Bpar}). For example, the symmetric contributions due to the parallel field threading the surface state and a possible out-of-plane component are not included in the model. 
Furthermore, the top and bottom surface states might have different $L_B$ due to a possible variation in chemical potential between top and bottom surfaces, which contribute separately to $\Delta \sigma_{xx}$. More details on both the in-plane and out-of-plane magnetoresistance are provided in \cite{MR_app}. 
The most important parameter for our analysis, however, is the magnetic field offset extracted from the fit.

The offset extracted from the fit is shown in Fig.~\ref{fig:inplane_results}(b). For low $I_\mathrm{DC}$ magnetic field shift increases, whereas for high $I_\mathrm{DC}$ the shift is suppressed.
We compare the extracted offset to the calculated Overhauser field for a single surface, shown in Fig.~\ref{fig:inplane_results}(b). To provide a range of expected $B_\mathrm{OH}$ values we assume $A_0 = 5 - 50$ $\mu$eV \cite{bozkurt2024entropy}, similar to the previous predictions based on Eqs. (\ref{eq:Boh}) and (\ref{eq:mbar}).
At very high values of the bias current, we expect additional physics to be active, which we discuss later. 

\begin{figure}
    \centering
    \includegraphics[width=\columnwidth]{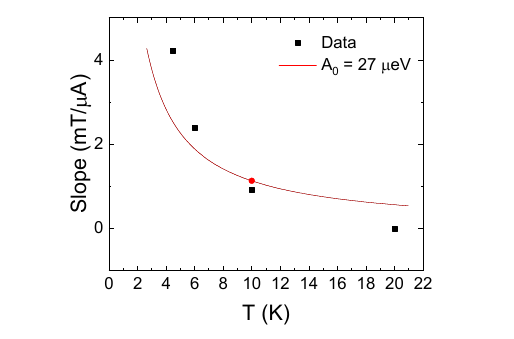}
    \caption{Temperature dependence of the low-current slope $\dd B_\mathrm{offs}/\dd I_\mathrm{DC}$, from fitting Eq. \ref{eq:Bpar}. We compare the data to the low-current slope from Eqs. (\ref{eq:Boh}) and (\ref{eq:mbar}) using a value of $A_0 = 27\ \mu$eV \cite{nisson2013nuclear}. See \cite{temp_app} for more details.}
    \label{fig:temp_dep_main}
\end{figure}

First, we focus on low $I_\mathrm{DC}$ and extract the nuclear coupling constant by comparing the data to Eqs. (\ref{eq:Boh}) and (\ref{eq:mbar}). 
The extracted value ($A_0\approx 46$ $\mu$eV for $-10$ $\mu$A$\leq I_\mathrm{DC}\leq 10$ $\mu$A) is of the same order of magnitude as the value ($A_0 = 27$ $\mu$eV) measured by Nisson \textit{et al.} using NMR at $T = 10$ K. Upon repeating the low-current measurements at $10$ K, we find closer correspondence with the value from Nisson \textit{et al.} \cite{temp_app}. We plot the temperature-dependence of the low-current slope $\dd B_\mathrm{offs}/\dd I_\mathrm{DC}$ in Fig. \ref{fig:temp_dep_main}, where the value $A_0 = 27\ \mu$eV under-predicts the data at low-temperatures and over-predicts the data at elevated temperatures. As $A_0$ is proportional to the electron density at the nuclear spin site \cite{dyakonov1984theory}, we can expect an enhancement of $A_0$ at low temperatures via the surface state penetration depth \cite{zhang2010first} compared to elevated temperatures. 
Combining these considerations, the results suggest that $B_\mathrm{offs}$ at low current can be attributed to the Overhauser field.

Beyond $|I_\mathrm{DC}| = 10$ $\mu$A, $B_\mathrm{offs}$ is reduced with increasing current.
Although $B_\mathrm{offs}$ falls within the limits of the expected  $B_\mathrm{OH}$, the non-monotonic relation between the extracted offset and $I_\mathrm{DC}$ suggests that additional effects are present.

One possibility is that Joule heating of the electron temperature affects the steady-state nuclear polarization. To illustrate this, we consider Eq.~(\ref{eq:mbar}), where the varying $I_\mathrm{DC}$ enters twice: via the bias voltage $V$ and the electron temperature $T$. From a rough comparison between bias- and temperature-dependent datasets we estimate that the electron temperature increases to approximately 20 K at $I_\mathrm{DC} = 100$ $\mu$A \cite{MR_app}. 
When measuring $B_\mathrm{offs}$ at $T = 20$ K \cite{temp_app}, we find that the shift is suppressed and vanishes. 
If the hyperfine interaction is invariant with temperature, the estimated Overhauser field is reduced from approximately 180 mT ($I_\mathrm{DC} = 100\ \mu$A, $T = 4.5$ K) to 45 mT ($I_\mathrm{DC} = 100\ \mu$A, $T = 20$ K). However, if the surface states are more delocalized at elevated electron temperatures, $B_\mathrm{OH}$ can decrease through $A_0$ because the hyperfine coupling strength is proportional to the electron density at the nucleus \cite{dyakonov1984theory}, reducing the Overhauser field below the modeled value.

Another possibility for the reduction of $B_\mathrm{offs}$ is the contribution from a second surface, whereas up to now, we have considered the Overhauser field from a single surface. Due to the helicity of the surface states, the expected nuclear polarization is opposite on top/bottom surfaces. The transport properties (e.g. mean free path, carrier density) can differ between the surfaces \cite{he2012highly}, which could lead to the equilibrium nuclear polarization on each surface scaling differently with applied bias voltage. The change in slope would occur when the nuclear polarization on one surface saturates, i.e. $(\ell_\mathrm{el}/L)\cdot eV/(2k_\mathrm{BT})\gtrsim 1$. At $V = 150$ mV ($I_\mathrm{DC} = 10$ $\mu$A and $T = 4.5$ K this would require $\ell_\mathrm{el}\gtrsim 0.3$ $\mu$m on one surface. However, fitting $B_\mathrm{offs}$ using two opposite Overhauser fields results in a low value of $A_0 \approx 3$ $\mu$eV. Again, Joule heating could provide the explanation for this low fit value: if the electron temperature is elevated, larger bias currents are required to achieve nuclear polarization in Eq. (\ref{eq:mbar}). This would reduce the slope $\dd B_\mathrm{offs}/\dd I$, which emerges as a reduction of the fit parameter setting this the slope, $A_0$ in Eq. (\ref{eq:Boh})

To constrain the fit of $B_\mathrm{offs}$ containing top- and bottom surface, the transport parameters from both surfaces can be obtained from a multi-band fit of the magnetoresistance.
Considering that our metallic leads mostly contact the top surface and that the linear Hall resistance is already well-explained by a single band Drude model \cite{MR_app}, we do not perform the multi-band fit. If these fit parameters would be available, the model covering the complete $I_\mathrm{DC}$ range would still require a detailed dependence of the electron temperature on bias current. Therefore, we limit our conclusion to the low-current regime, where we use a single-band model without Joule heating to find that $B_\mathrm{offs}$ is well-described by the Overhauser field.

Besides the Overhauser field, other mechanisms could enhance $B_\mathrm{offs}$. For example, the effective field resulting from exchange interactions between electron spins could add to the offset \cite{Bfields_app},  or additional magnetization could originate from localized unpaired electrons, due to lattice defects \cite{nisson2013nuclear, koumoulis2014understanding} Some additional discrepancy between our results and the expected Overhauser field could be explained by our assumption of $g = 30$ in Eq. (\ref{eq:Boh}) \cite{drath1967band}. Any change in $g$ would be proportional to a change in the calculated $A_0$.

\textit{Conclusion --} We searched for signatures of steady-state nuclear polarization in BST Hall bars using an in-plane magnetic field, perpendicular to the current. By applying a DC bias, we expected to polarize nuclear spins and thereby generate an Overhauser field, resulting in an offset in-plane magnetoresistance. 
At low currents, the extracted offset from our measurements reasonably matches the Overhauser field based on literature \cite{nisson2013nuclear}.
Although not exact, the results underline that our method, purely based on magnetotransport, can probe the interaction between electron spins in the topological surface state and nuclear spins.

Because the modeled Overhauser field by itself cannot explain the full current-dependence of $B_\mathrm{offs}$, the data presented in this article leaves several opportunities for future research.
The suppression of the extracted offset at high currents could be due to Joule heating suppressing the steady-state nuclear polarization, or due to both the top and bottom surface contributing to nuclear polarization. A complete fit would require (i) the relation between electron temperature and bias voltage and (ii) multi-band analysis of the transport data.
An additional contribution to the offset could result from electron spin-spin interactions, which is likely small compared to the Overhauser field, although its order of magnitude remains an open question to be resolved. 

Conducting a similar experiment in a three-dimensional topological insulator with spin-momentum-locked surface states, but with lower nuclear spin abundance, could further solidify this conclusion. One example is the topological crystalline insulator Pb$_{1-x}$Sn$_x$Te. Pb$_{1-x}$Sn$_x$Te hosts spin-momentum locked surface states as well \cite{xu2012observation}, and the nuclear spin in this material is reduced compared to BST (Pb has spin 1/2 at 22\% abundancy, Sn has spin 1/2 at 16\% abundancy, and Te has spin 1/2 at 7\% abundance \cite{schliemann2003electron}). Any Overhauser effect in Pb$_{1-x}$Sn$_x$Te is thus expected to be smaller than in \bsttight.

Further tests should include measuring the nuclear polarization timescale, which would require reducing the acquisition time of magnetotransport traces to timescales faster than the typical nuclear polarization dynamics. Similar to the effect measured by Jian \textit{et al.} in Bi(111) \cite{jiang2020dynamic}, the Overhauser field and consequential dephasing effects would persist after switching the measuring current from a high to a low value, until the nuclear spin polarization relaxes to the equilibrium (disordered) state. 
Finally, rotating the magnetic field in-plane (from perpendicular to collinear with the current) would shed further insight on the directionality of the shifted magnetoresistance, which can in turn be related to the nuclear polarization axis.

\textit{Acknowledgements} -- We thank Danielle Couger and Rodolfo Salas for fruitful discussions. 
S.K. deposited BST, fabricated devices, performed transport measurements, and wrote the manuscript with input from A.B. and I.A.. A.B. supervised the project. This research was supported by a Lockheed Martin Corporation Research Grant.\nocite{data}

\FloatBarrier
\bibliography{references}

\end{document}


\title{
Supplemental material to: \\Signature of current-induced nuclear spin polarization in (Bi$_{1-x}$Sb$_{x}$)$_2$Te$_3$
}
\author{Sofie Kölling, 
\.{I}nan\c{c} Adagideli, 
Alexander Brinkman}

\date{\today}
\renewcommand\thesection{\Alph{section}}
\newcommand{\supplementarysection}{%
  \setcounter{figure}{0}
  \let\oldthefigure\thefigure
  \renewcommand{\thefigure}{S\oldthefigure}
  }
\newcommand{\bst}{(Bi$_{1-x}$Sb$_x$)$_2$Te$_3$ }
\newcommand{\bsttight}{(Bi$_{1-x}$Sb$_x$)$_2$Te$_3$}

\maketitle

\supplementarysection
\section{Dynamic nuclear polarization}\label{sec:appDNP_DCBST}
\begin{figure}[h!]
    \centering
    \includegraphics{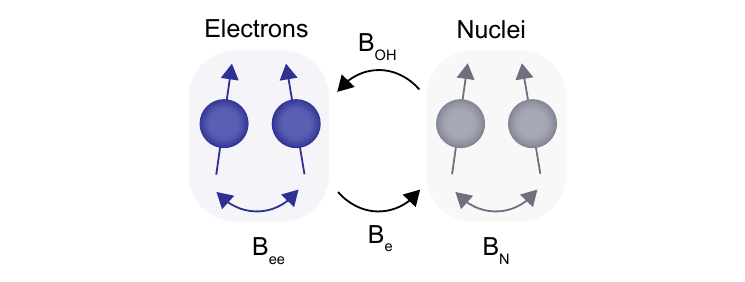}
    \caption{Spin-spin interactions in a topological surface state coupled to a nuclear spin subsystem, where we model the interactions as effective current-induced magnetic fields. The electron- and nuclear spin systems are coupled via hyperfine interaction; the effective field of an electron spin polarization acting on nuclear spins is $B_e$, and reversedly the effective field of a nuclear spin polarization acting on electrons is $B_\mathrm{OH}$. Nuclear spins interact via dipole-dipole interaction described by the effective field $B_N$, and electron spins via an effective interaction field $B_\mathrm{ee}$. The Oersted field is not included in this picture.  }
    \label{fig:Bfields}
\end{figure}
Having estimated the Overhauser field, we check whether the electron spin polarization is sufficient to achieve such nuclear spin polarization values. In order to achieve current-induced nuclear polarization in a topological surface state, the magnetic field generated by spin-polarized electrons ($B_e$) must exceed the dipole-dipole interaction field between nuclei ($B_N$). In systems where $B_e < B_N$, an external in-plane magnetic field is required for dynamic nuclear polarization \cite{dyakonov1984theory, salis2009signatures}.

The dipole-dipole interaction field between nuclear spins is given by \cite{dyakonov1984theory}
\begin{equation}
    B_N = \frac{\mu_0}{4\pi}\frac{\mu_I}{a^3},
\end{equation}
where $\mu_0$ is the vacuum permeability, $\mu_I = \mu_N I$ is the nuclear magnetic moment ($I$ is the nuclear spin and $\mu_N$ the nuclear magneton) and $a$ is the lattice constant.
As an upper limit for $B_L$ we consider two bismuth atoms (element with the strongest hyperfine interaction in BST) on the spacing in the BST lattice. Using $I = 9/2$ and $a = 4.3$ \AA\ \cite{mulder2022revisiting}, we find $B_N = 29$ $\mu$T.

The magnetic field generated by spin-polarized electrons, acting on nuclear spins, is calculated as \cite{fleisher1984optical} 
\begin{equation}
    B_e = n_e v_0 \frac{A_0 S_\mathrm{av}}{\mu_N}
\end{equation}
where $S_\mathrm{av}$ ($-1/2 \leq S_\mathrm{av} \leq 1/2$) is the average electron spin polarization, and $n_e$ is the electron density (per volume). The nonequilibrium spin accumulation in a topological surface state due to a charge current equals $\mathbf{S} = -\frac{\hbar}{2ev_F} \mathbf{j}\times \mathbf{\hat{z}}$ (in an ideal Dirac cone) \cite{kondou2016fermi}. For simplicity we assume $\mathbf{j} = j\mathbf{\hat{x}}$. The obtained value of $S$ gives the accumulated spin angular momentum per surface area. Note that hexagonal warping introduces an out-of-plane component in $\bm{S}_{av}$. However, this component would not introduce an offset when probing the in-plane magnetoresistance, so we only consider the in-plane spin polarization. To convert this to $S_\mathrm{av}$, we write
\begin{equation}
    S_\mathrm{av} =\frac{1}{2} \frac{j}{e v_F n_s}.
\end{equation}
For our films, typically $j = 1$ A/m, $n_s = 1\cdot 10^{13}$ cm$^2$ (see Supplementary section C), and $v_F = 4.7 \cdot 10^5$ m/s \cite{mulder2022spectroscopic}, leading to $S_\mathrm{av} = 1\cdot10^{-3}$. Assuming that the bulk is insulating and the current is concentrated on the top surface within 1 nm, we find $n_e = 1\cdot 10^{26}$ m$^{-3}$. Again using a range of $A_0 = 5-50$ $\mu$eV for BST, we find $B_e = 0.5 - 5$ mT. The estimated $B_e$ is an order of magnitude larger than $B_N$, so we could expect current-induced nuclear polarization in \bsttight.
\FloatBarrier
\clearpage
\section{Out-of-plane magnetoresistance}\label{sec:appMR_DCBST}
\begin{figure}[h!]
    \centering
    \includegraphics[width=\textwidth]{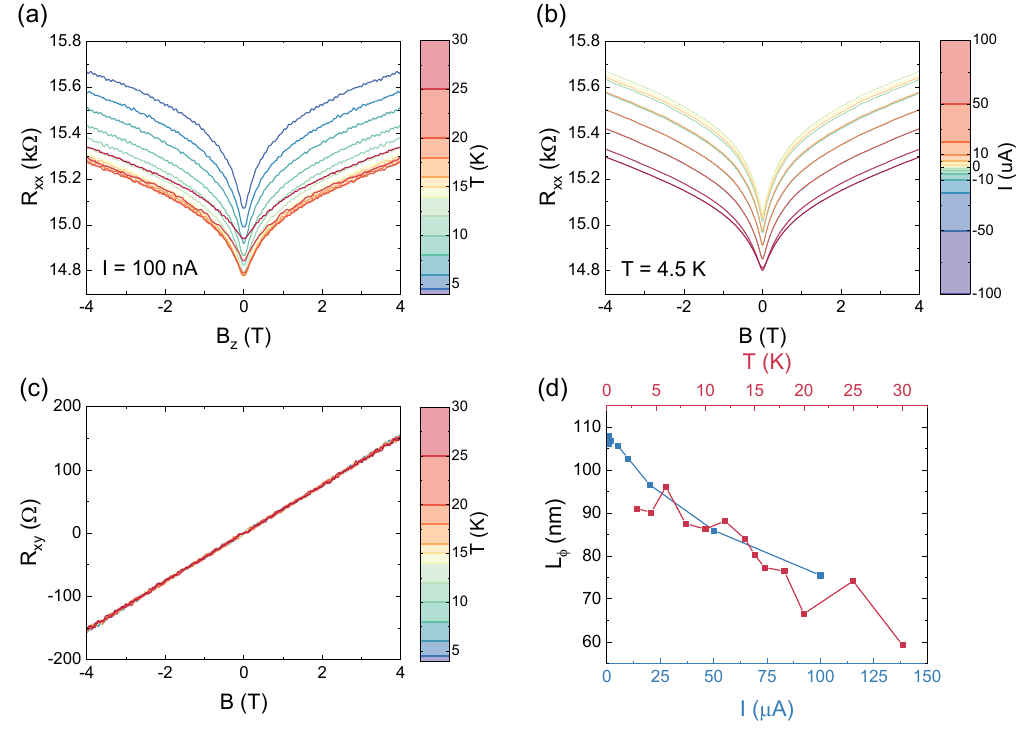}
    \caption{Longitudinal resistance in a (Bi$_{1-x}$Sb$_{x}$)$_2$Te$_3$ Hall bar as a function of magnetic field normal to the surface for a range of (a) temperatures and (b) bias currents (traces from positive and negative currents overlap). The data has been symmetrized in magnetic field. (c) $R_{xy}$ accompanying the data in (a). (d) Phase coherence length $L_\phi$, extracted by fitting the Hikami-Larkin-Nagaoka formula.}
    \label{fig:MR_oop}
\end{figure}

Figures \ref{fig:MR_oop}(a) and (b) show the magnetoresistance of the device studied in the main text with the magnetic field oriented out-of-plane and normal to the surface, at varying temperature and current bias.
We extract the phase coherence length $L_\phi$, used in Fig. \ref{fig:lengths} in the main text, by converting $R_{xx}$ to $\sigma_{xx}$ and fitting the Hikami-Larkin-Nagaoka (HLN) formula on the datasets of (a) and (b) up to $B = 2$ T \cite{hikami1980spin}. The HLN formula is given by
\begin{equation}\label{eq:hln}
    \sigma_{xx}(B) - \sigma_{xx}(0) = -\frac{\alpha e^2}{2\pi^2\hbar}\left(\ln \frac{B_\phi}{B} - \psi\left(\frac{1}{2} + \frac{B_\phi}{B} \right) \right).
\end{equation}
Here, $\alpha$ is a prefactor related to the number of parallel channels contributing to the quantum correction \cite{garate2012weak}, and $B_\phi = \hbar/4eL_\phi^2$.

The extracted phase coherence lengths from Figs.~\ref{fig:MR_oop}(a) and (b) are shown in Fig. \ref{fig:MR_oop}(c). Note that (a) and (b) were measured in different thermal cycles, which could explain the discrepancy between the low-temperature and low-bias $L_\phi$. The bias-dependent $L_\phi$ was measured in the same thermal cycle as the in-plane magnetoresistance of Figs. 1-2 in the main text. 
We also measured $R_{xy}$ in Fig.~\ref{fig:MR_oop}(c), and using a single-band Drude model we find $n_s = 1.6 \cdot 10^{13}$ cm$^{-2}$.

Dephasing due to an in-plane current-induced magnetic field, as discussed in the main text, suppresses the phase coherence time. This reduction of phase coherence emerges when measuring the magnetoresistance using an out-of-plane external magnetic field ($\bm{B}= B\hat{z}$) magnetic field as a broadening (suppression) of the WAL cusp. Experimentally, such a suppression of WAL due to an Overhauser field has been measured using an out-of-plane magnetic field in the Rashba system Bi(111) \cite{jiang2020dynamic}.

Although the two datasets from different thermal cycles cannot be compared directly, the similarity between the temperature- and bias-dependence could imply Joule heating of the electron temperature, as described by Abrahams and Anderson \cite{abrahams1980non, anderson1979possible}, and which has been studied more recently in magnetically doped topological insulators \cite{nandi2018logarithmic}. 
Joule heating elevates the electron temperature, which suppresses WAL as well. However, Joule heating and nuclear polarization effects could be distinguished if the nuclear polarization timescale is longer than the electron-phonon relaxation time. In this case, after removing the DC bias, the Overhauser field would persist, while the electron temperature equilibrates to the lattice temperature. 

Possibly, WAL broadening due to dephasing from an Overhauser field is present, but not observable on top of the heating effects in the out-of-plane magnetoresistance.
The typical width of the WAL cusp ($\sim$ 1 T) far exceeds $B_\mathrm{OH}$ ($\sim$ 10 mT), and consequently, the out-of-plane magnetic field dominates the dephasing. In more detail, the dephasing length $\Tilde{L}$ depends on both $L_\phi$ and $L_B$, of which the latter contains the Overhauser field. If we assume that current-induced in-plane magnetic field equals $B_\mathrm{OH}= 20$ mT (the external field is purely out-of-plane), $L_B \approx 1.7$ $\mu$m $\ll L_\phi$. Comparing these length scales, $L_\phi$ dominates dephasing in out-of-plane magnetoresistance, and the effect of an Overhauser field in this out-of-plane configuration is negligible.

\FloatBarrier
\clearpage
\section{Magnetic length and dephasing length}
\begin{figure}[h!]
    \centering
    \includegraphics{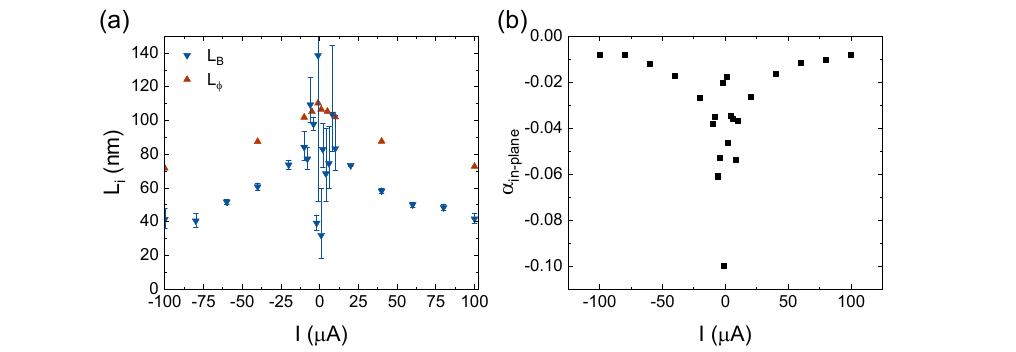}
    \caption{(a) Length scales extracted from magnetoresistance at $T = 4.5$ K. $L_\phi$ follows from an HLN fit on the out-of-plane magnetoresistance. $L_B$ is extracted by fitting Eq. (5) in the main text on the data also shown in Fig. 2(a), via $L_B = L_\phi B_c/B$ where $B = 0.5$ T. The error bars in $L_B$ denote the range of values obtained when analyzing multiple measured traces. (b) Fit parameter $\alpha$ obtained by fitting Eq. (5).}
    \label{fig:lengths}
\end{figure}
To apply Eq. (5) on in the in-plane configuration as done in the main text, we require $L_B$ to be the dominant dephasing length. 
$L_B$ and $L_\phi$ are compared in Fig.~\ref{fig:lengths}(a). For $L_\phi$ we refer to the HLN fit used to analyze the out-of-plane magnetoresistance, see Eq. (\ref{eq:hln}). At $T = 4.5$ K and low current, $L_\phi \approx 110$ nm. From the fit on Fig. 2(a) we obtain $B_c \approx 0.3$ T at $I_\mathrm{DC} = 1$ $\mu$A, resulting in $L_B \approx 60$ nm at $B = 0.5$ T. Note that this value was calculated using $B_\mathrm{ext} = 0.5$ T, whereas for consistency $B_\mathrm{eff}$ should be used. However, the contribution of $B_\mathrm{offs}$ is small on these field scales. In the in-plane magnetoresistance, $L_B<L_\phi$ at $B = 0.5$ T over the entire range of measured currents, so we indeed expect the Overhauser field (contained in $L_B$) to be observable in the magnetoresistance.

Figure \ref{fig:lengths}(b) shows the value of $\alpha$ required to fit Eq. (5). The value strongly deviates from $\alpha = 0.5$ predicted by the Adroguer \textit{et al.} model \cite{adroguer2012diffusion}, so a more detailed model is required to capture the full physics. Note that the magnitude of the in-plane magnetoresistance is sensitive to alignment errors of the external magnetic field. We minimized this error by rotating the sample in the external magnetic field until $R_{xy}$ reached a minimum.

\FloatBarrier
\clearpage
\section{Other current-induced magnetic fields}\label{sec:appBfields_DCBST}

\subsection{Electron-electron exchange field}
Apart from the Overhauser field, originating from a finite mean nuclear spin polarization, we can also consider the effects of a finite mean electron spin polarization acting back on electrons. This finite spin polarization is a direct consequence of spin-momentum locking in the topological surface state. By modeling the resulting effective magnetic field ($B_\mathrm{ee}$) as a Zeeman interaction, it adds to the Overhauser field and an external (in-plane) magnetic field.

Balram \textit{et al.} consider electron-electron interactions in a 1D helical edge state \cite{balram2019current}. They model the electron-electron interactions in a mean-field picture, using a contact interaction potential, as
$H_\mathrm{MF} = -\mathcal{J}_\mathrm{ee} \langle\bm{S}\rangle \cdot \bm{\sigma}/\hbar$, where $\langle\bm{S}\rangle$ is the expectation value of the electron spin polarization per unit length in the helical edge state. Rewriting this expression to an effective magnetic field ($H_\mathrm{MF} = g\mu_B \bm{B_\mathrm{ee}} \cdot \bm{\sigma}$) gives
\begin{equation}\label{eq:Hmf}
    B_\mathrm{ee} = -\frac{\mathcal{J}_\mathrm{ee}}{g \mu_B}  \frac{\langle \bm{S} \rangle}{\hbar}.
\end{equation}
Although time-reversal symmetry is conserved in the spin-momentum locked surface states, $B_\mathrm{ee}$ breaks time-reversal symmetry, as it considers the mean electron polarization as an external contribution acting on an electron (coupled via $\mathcal{J}_\mathrm{ee}$). Therefore, Eq.~(\ref{eq:Hmf}) is equivalent to an effective external magnetic field. 

The electron-electron field depends on the spin-spin coupling strength between electrons, which could be obtained from Fermi liquid theory \cite{lundgren2015landau}. We do not estimate an explicit value for $B_\mathrm{ee}$, as we expect this contribution to be small, as the system is in a non-magnetic state. The possibility of a magnetic instability due to ferromagnetic coupling between electron spins has been studied for the surface of a topological insulator, but the realistic interaction strength between electrons is too weak for such an instability to occur \cite{baum2012magnetic}.  A more accurate prediction of $B_{ee}$, obtained by considering spin-spin interactions in Fermi liquid theory \cite{lundgren2015landau}, requires further theoretical research.

\subsection{Oersted field}
Besides current-induced magnetic fields arising from finite mean nuclear- and electron spin polarizations, the moving charges on the topological surface state will generate a magnetic field as well, which is called the Oersted field.
We estimate an upper limit to the Oersted field by considering Ampère's law. The TI surface state is approximated by an infinite plane in $x-y$, with current along the $x$-direction. Then, the magnetic field is given by 
\begin{equation}
    \Vec{B} = \pm \mu_0 \frac{j}{2}\hat{y},
\end{equation}
where $j$ is the sheet current density and the change of sign corresponds to the field above or below the plane. Using $j = 1$ A/m results in $B = 0.6$ $\mu$T. This field is much smaller than the other magnetic fields calculated in this section, so we will not consider it in further analysis.
\clearpage

\section{Temperature dependence}
\begin{figure}[h!]
    \centering
    \includegraphics[width=\textwidth]{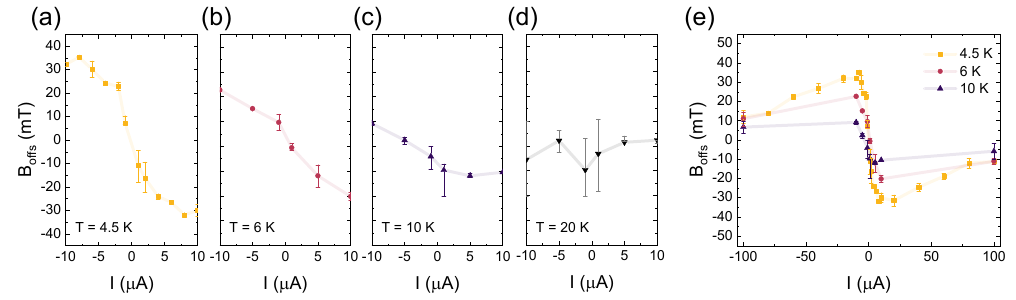}
    \caption{Temperature dependence of $B_\mathrm{offs}$, from fitting Eq. (5) for (a) 4.5 K, (b) 6 K, (c) 10 K and (d) 20 K, at low $I_\mathrm{DC}$. (e) $B_\mathrm{offs}$ for $-100\ \mu$A $\leq I_\mathrm{DC}\leq100\ \mu$A.}
    \label{fig:temp_dep}
\end{figure}

Figure \ref{fig:temp_dep} shows the data used to obtain the temperature dependent slope $\dd B_\mathrm{offs}/\dd I_\mathrm{DC}$ in Fig. 3 in the main text. By linearizing Eqs. (3) and (4), see the main text, the Overhauser field at low current (where $\frac{\ell_\mathrm{el}}{L}\frac{eV}{2k_\mathrm{B}T}\ll 1$) is described by
\begin{equation}
    B_\mathrm{OH} \approx [n] \frac{A_0 I}{g \mu_B} \frac{\ell_\mathrm{el}}{L}\frac{eV}{2k_\mathrm{B}T}. 
\end{equation}
By substituting $V = R_{xx}I_\mathrm{DC}$ (where the current, $I_\mathrm{DC}$, should not be confused with the nuclear polarization, $I$) we find
\begin{equation}\label{eq:slope}
    \dv{B_\mathrm{OH}}{I_\mathrm{DC}} \approx [n] \frac{A_0 I}{g \mu_B} \frac{\ell_\mathrm{el}}{L}\frac{eR_{xx}}{2k_\mathrm{B}T}.
\end{equation}
To obtain Fig. 3 in the main text, we use $L = 60 \ \mu$m, $\ell_\mathrm{el} = 10$ nm, $R_{xx} = 15$ k$\Omega$, $[n]I = \frac{1}{2}\cdot44.2$ with an antimony ratio of $x = 0.52$ in \bst \cite{schliemann2003electron}, $A_0 = 27$ $\mu$eV \cite{nisson2013nuclear} and $g = 30$ \cite{drath1967band}.

\clearpage
\FloatBarrier
\printbibliography